\documentclass[preprint]{JHEP3} 

\usepackage{epsfig,multicol}

\newcommand\fverb{\setbox\pippobox=\hbox\bgroup\verb}
\newcommand\fverbdo{\egroup\medskip\noindent%
            \fbox{\unhbox\pippobox}\ }
\newcommand\fverbit{\egroup\item[\fbox{\unhbox\pippobox}]}
\newbox\pippobox

\newcommand{\ovl}{\overline}
\newcommand{\be}{\begin{equation}}
\newcommand{\ee}{\end{equation}}
\newcommand{\ba}{\begin{eqnarray}}
\newcommand{\ea}{\end{eqnarray}}
\newcommand{\bt}{\begin{table}}
\newcommand{\et}{\end{table}}
\newcommand{\brt}{\begin{ruledtabular}}
\newcommand{\ert}{\end{ruledtabular}}
\newcommand{\btu}{\begin{tabular}}
\newcommand{\etu}{\end{tabular}}
\newcommand{\la}{\langle}
\newcommand{\ra}{\rangle}
\newcommand{\non}{\nonumber}

\def\vev#1{\langle{#1}\rangle}

\title{Recursion Relations for Tree Amplitudes in Super Gauge Theories}

\author{Mingxing Luo and Congkao Wen\\
Zhejiang Institute of Modern Physics, Department of Physics \\
Zhejiang University, Hangzhou, Zhejiang 310027, P R China \\
    E-mail: \email{luo@zimp.zju.edu.cn, wenagua@sina.com}}
\date{\today}        

\preprint{\hepth{0501121}}  

\abstract{Using newly proposed recursion relations in \cite{bcf,bcfw},
compact formulas are obtained for tree-level amplitudes of $g_1^- g_2^- g_3^- g_4^+ g_5^+ \cdots g_n^+$.
We then make an extension of these recursion relations to include fermions of multi-flavors,
from which MHV and $\overline{\rm MHV}$ amplitudes are reproduced. 
We also calculate non-MHV amplitudes of processes with two fermions and four gluons.
Results thus obtained are equivalent to those obtained by extended CSW prescriptions,
and those by conventional field theory calculations. 
}


\begin{document}


\section{Introduction}

In perturbative gauge theories, extremely simple results usually emerge after tremendously complicated calculations 
by using conventional field theory methods.
For example, the Parke-Taylor formula for maximal helicity violating (MHV) amplitudes
can be written within one line, which summarizes numerous ordinary Feynman diagrams \cite{PT}. 

Recently, a deep relation was pointed out between $N=4$ super gauge theory and 
one type B topological string theory \cite{witten}, by re-expressing
super gauge theory scattering amplitudes in the language of twistor theories \cite{penrose}.
Taking advantage of insights thus gained and by a careful analysis of known helicity amplitudes, 
Cachazo, Svrcek, and Witten (CSW) \cite{csw} proposed a novel
prescription to calculate tree level amplitudes, 
which uses MHV amplitudes as vertices to construct all other amplitudes,
thus initiated the paradigm of MHV Feynman diagrams. 
The efficiency of the method is phenomenal and its validity has
been checked by various tree level calculations \cite{wuzhu,khoze,ggk,kosower,wuzhu2,bbk,wuzhu3,gk,zhu}. 
In \cite{bst}, the method was extended to calculate one-loop MHV amplitudes
and correct results were reproduced in the large $N_c$ limit.
It turns out that the paradigm is independent of the large $N_c$ approximation,
at least to one-loop \cite{lw1, lw2}. 
The twistor-space structure of one-loop amplitudes are further studied in \cite{csw2,csw3,bern10,cachazo10}.
On the other hand, tree-level
amplitudes were also obtained from connected curves in twistor string theories \cite{rsv1,rsv2,rsv3}.

However, the number of MHV diagrams grows rapidly when one includes more external particles,
though real amplitudes may well be much simpler than those suggested by CSW rules.
For example, an extremely simple result was obtained for an amplitude of eight gluons in \cite{rsv4},
by dissolving $N=4$ loop amplitudes into tree ones.
Related,
Britto, Cachazo and Feng (BCF) proposed a new set of recursion relations to calculate tree amplitudes \cite{bcf},
based upon analysis of one-loop amplitudes and infrared relations.
The new method has reproduced known results up to seven external gluons, and
can easily be applied to processes with any reasonable number of external gluons.
Soon after, the BCF prescription was proved alternatively by using basic facts of tree diagrams 
with some help from MHV Feynman diagrams \cite{bcfw} (BCFW).
In \cite{bcf}, the recursion relations used two adjacent gluons of opposite helicity as reference gluons.
However, it has been shown in \cite{bcfw} that reference gluons do not have to be adjacent and
they can also be of the same helicity.
Related works can be found in \cite{bk1,bk2}.

As a concrete application of the BCF prescription, 
we will in this paper calculate tree-level amplitudes of $g_1^- g_2^- g_3^- g_4^+ g_5^+ \cdots g_n^+$.
As expected, extremely compact formulas are obtained,
which will be presented in section 2 after a brief review of the BCF rules.
These recursion relations are then extended to include fermions of multi-flavors,
in the context of super gauge theories,
from which MHV and $\overline{\rm MHV}$ amplitudes are reproduced correctly (section 3). 
In section 4, we will calculate non-MHV amplitudes of processes with two fermions and four gluons,
by using of both our extensions and extended CSW prescriptions \cite{gk}.
Results obtained by these two methods are equivalent to each other,
and equivalent to those given in \cite{mp} by conventional field theory calculations.
These results provide strong evidences to support both the CSW and the BCF/BCFW prescriptions,
as well as the present extension.

\section{Review of the BCF/BCFW Approach and a Compact formula}
In four dimensional space-time, a four-vector $p_\mu$ can always be
expressed as a bispinor $p_{a\dot{a}}=p_\mu \sigma^\mu_{a\dot{a}}$,
and a null vector can be factorized $p_{a\dot{a}}=\lambda_a \tilde{\lambda}_{\dot{a}}$ 
in terms of spinors $\lambda_a$, $\tilde{\lambda}_{\dot{a}}$ of positive and negative chirality. 
Spinor products are defined to be
$\langle\lambda_1,\lambda_2\rangle=\epsilon_{ab}\lambda^a_1\lambda^b_2$
and $\langle\tilde{\lambda}_1,\tilde{\lambda}_2\rangle=
\epsilon_{\dot{a}\dot{b}}\tilde{\lambda}^{\dot{a}}_1\tilde{\lambda}^{\dot{b}}_2$,
which are usually abbreviated as $\langle 1,2 \rangle$ and $[1,2]$.
For two null vectors $p_{a\dot{a}}=\lambda_a \tilde{\lambda}_{\dot{a}}$ 
and $q_{a\dot{a}}=\mu_a \tilde{\mu}_{\dot{a}}$, 
one has $2p\cdot q= \langle \lambda, \mu \rangle [\tilde{\lambda}, \tilde{\mu}]$.

Now we give a brief review of the BCF/BCFW approach.
Take an $n$-gluon tree-level amplitude of any helicity configuration. 
As amplitudes of gluons with one helicity vanish,
we can always arrange gluons such that the $(n-1)$-th gluon has negative helicity
and the $n$-th gluon has positive helicity. These two lines will be taken as reference lines.
Labeling external particles by $i$, the following recursion relation was claimed in \cite{bcf}:
\ba
A_n(1,2,\ldots , (n-1)^-,n^+) & =& \sum_{i=1}^{n-3}\sum_{h=+,-} 
\left( A_{i+2}(\hat{n},1,2,\ldots, i,-\hat{P}^h_{n,i} ) \right. \label{eq2.1}  \\
&& \times {1\over P^2_{n,i}} \left. A_{n-i}(+\hat{P}^{-h}_{n,i}, i+1,\ldots , n-2, {\hat{n-1} } ) \right) \nonumber
\ea
where
\ba
P_{n,i} & = & p_n+p_1+\ldots + p_i, \nonumber \\ 
\hat{P}_{n,i} & =& P_{n,i} +{P_{n,i}^2\over \langle n-1|P_{n,i}|n]} \lambda_{n-1} \tilde\lambda_{n}, \nonumber\\ 
\hat p_{n-1} & = & p_{n-1} -{P_{n,i}^2\over \langle n-1|P_{n,i}|n]} \lambda_{n-1}\tilde\lambda_{n} , \\
\hat p_{n} & = & p_{n} +{P_{n,i}^2\over\langle n-1|P_{n,i}|n]} \lambda_{n-1} \tilde\lambda_{n}. \nonumber
\ea
The formula has a natural meaning in $(--++)$ signature and can be pictorially represented,
as shown in Figure 1.
But the recursion relations do not depend the signature and we will take it to be $(+---)$. 
Notice that ${\hat P}_{n,i}^2 = {\hat p}^2_n = {\hat p}^2_{n-1} = 0$, 
so each tree-level amplitude in eq (\ref{eq2.1}) has all external gluons on-shell. 
Still, energy-momentum conservation is preserved.
In \cite{bcfw}, 
it has been shown that reference gluons do not have to be adjacent and they can be of the same helicity.
As we shall see late, one could get equivalent but more compact results if reference particles are of the same helicity,
or the other way around.
\begin{figure}[h]
\begin{center}
\leavevmode
{\epsfxsize=3.5truein \epsfbox{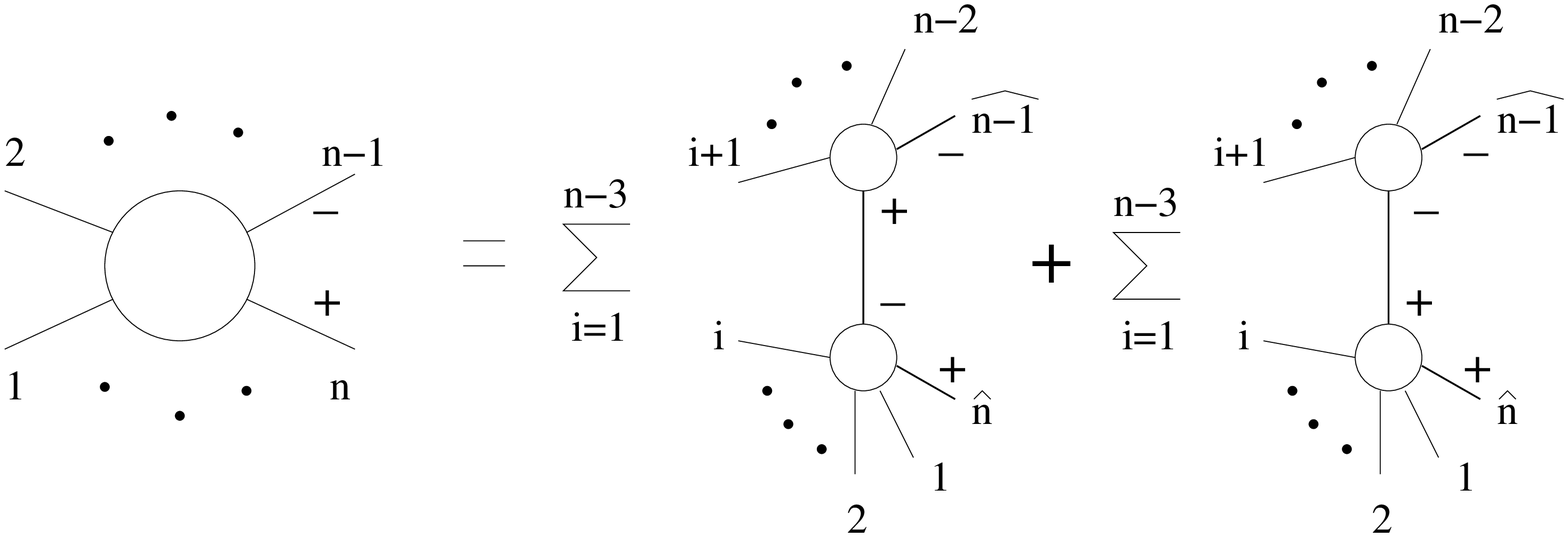}}
\end{center}
\caption{Pictorial representation of the recursion relation (2.1). 
Note that the difference between the terms in the two
sums is just the helicity assignment of the internal line. \cite{bcf}}
\end{figure}

To streamline notations, it is expedient to define 
\be
\begin{array}{rcl}
K_i^{[r]} &\equiv& p_i+p_{i+1}+\cdots+p_{i+r-1} \\ 
t_i^{[r]} &\equiv& \left( K_i^{[r]} \right)^2   \\
\langle i|\sum_r  p_r |j] &\equiv& \sum_r \langle i~r\rangle [r~j]  \\
\langle i|(\sum_r p_r )(\sum_s p_s )|j \rangle &\equiv& \sum_r\sum_s \vev{i~r}[r~s]\vev{s~j} 
\end{array}
\ee
And the following relations will be useful to make simplifications:
\be
\langle A | \hat{P}_{n,i} \rangle  = -{1\over \omega} \langle A | P_{n,i} |n], \ \ \ \
\left[ \right. \hat{P}_{n,i} | B ]  = -{1\over\bar{\omega}} \langle n-1 | P_{n,i} | B ]
\ee
Factors $\omega$ and $\bar{\omega}$ always show up in final results
in the combination of $\omega\bar{\omega}$, which is equal to $\la n-1 | P_{n,i} | n\ra$.
This completes the BCF/BCFW prescription.

Now, we use these rules to calculate tree-level amplitudes of $g_1^- g_2^- g_3^- g_4^+ g_5^+ \cdots g_n^+$.
After some experimentation, the following general result for arbitrary $n$ can be guessed
\ba
A(1^- 2^- 3^- 4^+ \cdots n^+) 
&=& {\la 1 | K_2^{[2]} |4]^3 \over t_2^{[3]} [2\ 3] [3\ 4] \la 5|K_3^{[2]} |2] \prod_{i=5}^n \la i\ i+1 \ra } \non \\
&-& \sum_{i=2}^{n-5} {\la 3 | K_4^{[i]} K_{4+i}^{[n-3-i]} |1\ra^3 \over 
t_3^{[i+1]} t_2^{[i+2]} \la i+3|K_3^{[i]} |2] \la i+4| K_3^{[i+1]}|2]}
{\la i+3 \ i+4 \ra \over \prod_{j=3}^n \la j\ j+1 \ra} \non\\
&+& {\la 3 | K_1^{[2]} |n ]^3 \over 
t_{n}^{[3]} [n\ 1] [1\ 2] \la n-1| K_n^{[2]} |2] \prod_{j=3}^{n-2} \la j\ j+1 \ra}
\ea 
Its validity can be easily proved by induction with the help of BCF rules.
On the other hand, one gets the above result directly if $g_1^-, \ g_2^-$ are chosen as reference lines.
These results are much compact than those given in \cite{kosower,kosower1}.

\section{Extensions to Include Fermions, MHV and $\ovl{\rm MHV}$ Amplitudes}
The BCF rules can be naturally extended to include fermions.
Specifically, one chooses any two external lines, either gluons or fermions, as reference lines;
then shifts relevant momenta exactly in the same manner as those in Eq. (2.2) and finally,
combines sub-amplitudes of less numbers of external lines together in the same manner as that in Eq. (2.1).
Of course, there could now be diagrams in which sub-diagrams are linked by internal fermionic lines,
but their propagators are the same as those of gluons.
For processes with different flavors of fermions, 
one should properly exclude diagrams in which fermions would be forced to change flavors in sub-diagrams.
For example, in a process of four fermions of two types, $(\Lambda_1^-,\Lambda_2^+,\eta_3^-,\eta_4^+)$,
one should exclude $u$-channel diagrams.
These extensions can be justified by similar reasonings as those in either \cite{bcf} or \cite{bcfw},
and by careful analysis of ordinary Feynman diagrams.

But there are some qualifications. 
Specifically, one can take neither two adjacent fermion lines of the same type and opposite helicities,
nor two adjacent fermion lines of different types and the same helicity,
nor one fermion and an adjacent gluon of the same helicity, as reference lines.
Otherwise one runs into inconsistencies.
Actually, the rational function $A(z)$ defined in Eq (2.3) of \cite{bcfw} 
would not vanish as $z \rightarrow \infty$ with such a choice, which spoils the recursion relations.
We shall come back to this later.

Now we prove by induction that 
these new rules generate correct MHV and $\ovl{\rm MHV}$ amplitudes with two or four fermions.
As reference lines can be chosen as wish, there are numerous but equivalent proofs.
Details of each proof also depend on relative positions between fermions and gluons.
One has to prove each case by exhausting all possibilities.
However, all proofs are similar and a full list would be tedious and dull.
We will instead give below three representative cases for illustration.
Other cases can be worked out accordingly.

The first example has the following structure,
$(g_1^-,g_2^+,\cdots,\Lambda_i^-,\cdots,\Lambda_j^+,\cdots,g_n^+)$ ($i>3$),
which has two fermions.
Assume generalized Parke-Taylor formulas for MHV amplitudes with fermions of $i<n$ external particles
and pick reference lines to be $g_1^-,\ g_2^+$, we have
\ba
\frac{\langle \widehat{1}\ i \rangle^3 \langle \widehat{1}\ j \rangle}
{\langle \widehat{1}\ \widehat{p} \rangle \langle \widehat{p}\ 4 \rangle 
\langle n\ \widehat{1} \rangle
\prod_{m=4}^{n-1}\langle m\ m+1 \rangle}
\times \frac{1}{t_2^{[2]}}
\times \frac{[\widehat{2}\ 3]^3}{[3\ \widehat{p}][\widehat{p}\ \widehat{2}]}
=\frac{\langle 1\ i \rangle^3 \langle 1\ j \rangle}{\prod_{m=1}^n \langle m\ m+1 \rangle}
\ea
which is just the generalized Parke-Taylor formula for MHV amplitudes with fermions of $n$ external particles.
The second example has four fermions of the same flavor and the following structure,
$(\Lambda_1^-,\Lambda_i^+,\Lambda_j^-,\Lambda_k^+)$ $(i>2, \ j>i+1)$ 
and the reference lines will be taken to be $\Lambda_1^-,\ \Lambda_i^+$.
It turns out that there are two terms. The first one is
\ba
A_1&=&
\frac{[i+1\ \widehat{i}]^3[i+1\ \widehat{p}]}{[i+1\ \widehat{p}][\widehat{p}\ \widehat{i}][\widehat{i}\ i+1]}
\times {1 \over t_i^{[2]}} \non\\
&& \times \frac{\langle \widehat{1}\ j \rangle ^3\langle \widehat{p}\ k\rangle}
{\langle \widehat{p},i+2 \rangle 
 \langle n\ \widehat{1} \rangle \langle \widehat{1}\ 2 \ra
 \langle i-1\ \widehat{p}\rangle 
\prod_{m=i+2}^{n-1} \langle m\ m+1 \rangle 
\prod_{r=2}^{i-2} \langle r\ r+1 \rangle }
 \non \\
&=&
\frac{\langle k\ i+1 \rangle \langle i-1\ i\rangle}{\langle k\ i \rangle \langle i-1\ i+1 \rangle}
\frac{\langle 1\ j \rangle ^3 \langle i\ k \rangle}{\prod_{m=1}^n \langle m\ m+1 \rangle} 
\ea
and the second one is
\ba
A_2&=&
{[i-1\ \widehat{i}]^3[i-1\ \widehat{p}] \over [i-1\ \widehat{i}][\widehat{i}\ \widehat{p}][\widehat{p}\ i-1]}
\times {1 \over t_{i-1}^{[2]}} \non\\
&& \times \frac{\langle \widehat{1}\ j  \rangle^3}{\langle \widehat{1}\ 2 \rangle
\langle i-2\ \widehat{p} \rangle \langle \widehat{p}\ i+1 \rangle 
\prod_{m=2}^{i-3} \langle m\  m+1 \rangle 
\prod_{r=i+1}^n \langle r\ r+1 \rangle} \non \\
&=&
\frac{\langle k\ i-1\rangle \langle i\ i+1\rangle}{\langle k\ i \rangle \langle i-1\ i+1 \rangle}
\frac{\langle 1\ j\rangle ^3 \langle i\ k \rangle}{\prod_{m=1}^n \langle m\ m+1\rangle} 
\ea
Using Schouten's identity 
\be
\langle i\ i+1 \rangle \langle k\ i-1 \rangle +\langle i-1\ i \rangle \langle k\ i+1 \rangle 
=\langle i\ k \rangle \langle i+1\ i-1 \rangle
\ee
to combine them together, we have
\ba
A(\Lambda_1^- \Lambda_i^+ \Lambda_j^- \Lambda_k^+)
= \frac{\langle 1\ j\rangle ^3 \langle i\ k \rangle}{\prod_{m=1}^n \langle m\ m+1\rangle}
\ea
which is the generalized Parke-Taylor formula for MHV amplitudes with four fermions of the same type.
The third example has four fermions of two flavors and the following structure,
$(\Lambda_1^-,\Lambda_i^+,\eta_j^-,\eta_k^+)$, if the reference lines are chosen to be $\Lambda_1^-$ and $g_2^+$,
one then has inductively
\ba
A(\Lambda_1^- \Lambda_i^+ \eta_j^- \eta_k^+) 
&=&\frac{[\hat{2}\ 3]^3}{[3\ \hat{p}][\hat{p}\ \hat{2}]} 
\times \frac{1}{t_2^{[2]}} \times
\frac{\langle \hat{1}\ j\rangle ^2\langle i\ j\rangle 
\langle \hat{1}\ k\rangle}{\langle \hat{1}\ \hat{p}\rangle 
\langle \hat{p}\ 4\rangle \langle n\ \hat{1} \rangle \prod_{m=4}^{n-1} \langle m\ m+1\rangle} \nonumber\\
&=&\frac{\langle 1\ j\rangle ^2 \langle i\ j \rangle \langle 1\ k \rangle }{\prod_{m=1}^n \langle m\ m+1\rangle}
\ea
which is the generalized Parke-Taylor formula for MHV amplitudes with four fermions of two flavors.
Thus complete inductions for these three cases.

As indicated above, other cases can be worked out similarly.
Almost identical reasonings lead to the same conclusion for $\ovl{\rm MHV}$ amplitudes.
In a sense, these results can be anticipated.
As pointed in \cite{bcf}, the BCF rules yield the correct MHV amplitudes of pure gluons.
On the other hand, $\ovl{\rm MHV}$ amplitudes of pure gluons can be also obtained inductively, 
by following the same steps.
From them, MHV and $\ovl{\rm MHV}$ amplitudes with two or four fermions can be obtained via Ward identities.
But proofs outlined above are welcome checks for self-consistency of our extensions.

\section{Scattering Amplitudes of $\Lambda_1^+ \Lambda_2^- g_3 g_4 g_5 g_6$}
Now we apply the rules in the previous section to calculate non-MHV amplitudes 
for processes with two fermions and four gluons.
Paralleling results in \cite{mp}, we calculate the following six amplitudes.

Choosing $p_5$ and $p_6$ as reference momenta, we have
\ba 
A(\Lambda_1^+ \Lambda_2^- g_3^- g_4^+ g_5^- g_6^+)
&=& - {\la 3 \ 2 \ra^2 \la 3 \ 1 \ra [4\ 6]^4 \over 
t_1^{[3]} \la 1 \ 2 \ra [4 \ 5] [5\ 6] \la 3 | 4+5 | 6] \la 1 | 5+6 | 4 ]}  \non \\
&& - {\la 1 \ 5 \ra [2\ 4] \la 5| 1+6 |4 ]^3 \over 
t_2^{[3]} \la 1 \ 6 \ra \la 5\ 6 \ra [2 \ 3] [3\ 4] \la 1 | 5+6 | 4] \la 5 | 1+6 | 2 ]}  \\
&& - {\la 3 \ 5 \ra^4 [6\ 1]^2 [6\ 2] \over 
t_3^{[3]} \la 3 \ 4 \ra \la 4 \ 5 \ra [1\ 2] \la 3 | 1+2 | 6] \la 5 | 1+6 | 2 ]}  \non
\ea
Choosing $p_4$ and $p_5$ as reference momenta, we have
\ba 
A(\Lambda_1^+ \Lambda_2^- g_3^- g_4^- g_5^+ g_6^+)
&=& -  {\la 4 | 5+6 | 1 ]^2 \la 4| 5+6 |2] \over 
t_4^{[3]} [1 \ 2] [2\ 3] \la 4 \ 5\ra \la 5\ 6\ra [3 | 4+5 | 6\ra }  \non \\
&& + {\la 2 | 3+4 | 5 ]^2 \la 1| 3+4 |5] \over 
t_3^{[3]} \la 1 \ 2 \ra \la 6\ 1 \ra [3 \ 4] [4\ 5] [3 | 4+5 | 6 \ra }  
\ea
Choosing $p_5$ and $p_6$ as reference momenta, we have
\ba 
A(\Lambda_1^+ \Lambda_2^- g_3^+ g_4^- g_5^- g_6^+)
&=& - {\la 1\ 5 \ra  \la 5 |1+6| 3]^3 \over 
t_2^{[3]} \la 1 \ 6 \ra [3 \ 4] \la 5\ 6 \ra [4 |5+6| 1 \ra \la 5 | 1+6 | 2 ]}  \non \\
&& + {\la 4 \ 5 \ra^3 [6\ 1]^2 [6\ 2 ] \over 
t_3^{[3]} \la 3 \ 4 \ra [1\ 2] \la 3 | 4+5 | 6] \la 5 | 1+6 | 2 ]}  \\
&& + {\la 1| 4+5 |6] \la 2 | 4+5 |6]^3 \over 
t_4^{[3]} \la 1 \ 2 \ra \la 2 \ 3 \ra [4\ 5] [5\ 6] \la 3 | 4+5 | 6] \la 1 | 5+6 | 4 ]}  \non
\ea
Choosing $p_3$ and $p_4$ as reference momenta, we have
\ba 
A(\Lambda_1^+ \Lambda_2^- g_3^- g_4^+ g_5^+ g_6^-)
&=& -  {\la 3 |4+5| 1]^3 \la 3 |4+5 |2] \over 
t_3^{[3]} [6\ 1] [1\ 2] \la 3 \ 4 \ra \la 4 \ 5\ra \la 3 | 4+5 | 6] [ 2 | 3+4 | 5 \rangle }  \non \\
&& + {\la 1 \ 3 \ra \la 2 \ 3 \ra^2 [4\ 5]^3 \over 
t_1^{[3]} \la 1 \ 2 \ra [5\ 6] \la 1 | 2+3 | 4] \la 3 | 1+2 | 6 ]}  \\
&& + {[2\ 4] \la 6| 2+3 |4]^3 \over 
t_2^{[3]} [2 \ 3] \la 5 \ 6 \ra [3\ 4] \la 1 | 2+3 | 4] [2 | 3+4 | 5 \ra }  \non
\ea
Choosing $p_2$ and $p_3$ as reference momenta, we have
\ba 
A(\Lambda_1^+ \Lambda_2^- g_3^+ g_4^- g_5^+ g_6^-)
&=&-  {[1 \ 3]^3 \la 4 \ 6\ra^4 \over 
t_4^{[3]} [1 \ 2] \la 4 \ 5\ra \la5\ 6\ra \la 6 | 1+2 | 3] [1 | 2+3 | 4\ra}  \non \\
&& + {\la 6 \ 2 \ra^3 [3\ 5]^4 \over 
t_3^{[3]} \la 1 \ 2 \ra [3 \ 4] [4\ 5] \la 6 | 5+4 | 3] \la 2 | 3+4 | 5 ]}  \\
&& + {\la 4 \ 2 \ra^3 [1\ 5]^3 \la 4| 1+6 |5 ] \over 
t_2^{[3]} \la 2 \ 3 \ra \la 3 \ 4 \ra [5\ 6] [6\ 1] \la 2 | 3+4 | 5] \la 4 | 3+2 | 1 ]}  \non
\ea
Choosing $p_2$ and $p_3$ as reference momenta, we have
\ba 
A(\Lambda_1^+ \Lambda_2^- g_3^+ g_4^+ g_5^- g_6^-)
&=&   { \la 2| 3+4 |1]^3 \over 
\la 2 \ 3\ra \la3 \ 4\ra [5\ 6] [6\ 1] \la 2 | 3+4 | 5] [1 | 2+3 | 4 \ra}  \non \\
&& + {[3 \ 4]^3  \la 6 \ 2 \ra^3 \over 
t_3^{[3]} \la 1 \ 2 \ra [4\ 5] \la 6 | 5+4 | 3] \la 2 | 3+4 | 5 ]}  \\
&& - {[1\ 3]^3 \la 5 \ 6\ra^3 \over 
t_1^{[3]} [1\ 2] \la 4\ 5 \ra \la 6 | 1+2 | 3] [1 | 2+3 | 4\ra }  \non
\ea
The first term in Eq (4.6) seems to be a little out of place, which carries no $t^{[3]}$ factor.
It turns out that this term can be merged into the other two by choosing different reference lines.
Specifically, by choosing $p_3$ and $p_4$ as reference momenta, 
(note that lines 3 and 4 have the same helicity),
we can get an equivalent but more compact result
\ba
A(\Lambda_1^+ \Lambda_2^- g_3^+ g_4^+ g_5^- g_6^-)
&=&   { \la 2| 5+6 |4]^3 \over 
t_1^{[3]} \la 1\ 2 \ra \la 2\ 3 \ra [4\ 5] [5\ 6] \la 3 | 4+5 | 6] }  \non\\
&& - { \la 5| 4+3 |1]^3 \over 
t_3^{[3]} [6\ 1] [1\ 2] \la 3 \ 4 \ra \la4\ 5\ra \la 3 | 4+5 | 6]}
\ea
These results are quite simple and compact.
We would like to emphasis that forms of these results depend on the choice of reference lines,
though all expressions are equivalent.
There could be simpler expressions for these amplitudes if one chooses more clever reference lines.

Results in Eq (4.1--7) agree fully with those in \cite{mp} which were obtained by conventional field theory,
up to a possible overall sign,\footnote{
In previous versions of this paper, it was claimed that our results do not agree with those in \cite{mp}.
The mismatch was due to the fact that spinor products used in \cite{mp} and those we are using differ by a minus sign.
Fortunately, or unfortunately, this minus sign cancels out in cases of pure gluons,
which prevented it to be located for quite a while. 
We thank L. Dixon and C. Berger for assuring us that Eq (4.5--7) are equivalent to the corresponding ones in \cite{mp},
so the authors were encouraged enough to finally resolve the issue by digging into the computer codes again.}
but assume extremely simple and compact forms.
We have also computed these six amplitudes by using extended CSW prescriptions \cite{gk},
and found them to be equivalent (usually in different forms) to the ones shown above.
It can also be easily checked that our results satisfy all supersymmetric Ward identities.

\section{Conclusion}
In this paper, we have calculated tree-level amplitudes of $g_1^- g_2^- g_3^- g_4^+ g_5^+ \cdots g_n^+$
in an application of the BCF/BFCW rules.
As expected, extremely compact formulas were obtained.
The BCF/BCFW recursion relations were then extended to include fermions of multi-flavors,
from which MHV and $\overline{\rm MHV}$ amplitudes are reproduced correctly. 
We have also calculated non-MHV amplitudes of processes with two fermions and four gluons.
Results thus obtained are equivalent to those obtained by using extended CSW prescriptions,
and those by conventional field theory calculations.
These results provide more strong evidences to support both the CSW and the BCF/BCFW prescriptions,
as well as present extensions to include fermions.
As a distinctive feature, our results assume extremely simple and compact forms
and the method is elegant enough so, many difficult amplitudes can be calculated with relative easy.

Before closing, let us try to justify the extended rules by following the BCFW proof.
The BCFW proof has three important ingredients: 
the construction of a rational function $A(z)$ of $z$, 
the assertion that $A(z)$ has only simple $z$-poles, and that $A(z)$ vanishes as $z \rightarrow \infty$.
From which one has
\be
A(z) = \sum_{i,j} {c_{ij} \over z - z_{ij}}
\ee
where $C_{ij}$ are independent of $z$.
Thus, $z$-pole positions and residues determine the physical amplitude $A(0)$ uniquely.
When fermions are included, 
the first two ingredients can easily be established by simply repeating reasonings of \cite{bcfw}.
The importance of a vanishing $A(z)$ as $z \rightarrow \infty$ is obvious. Otherwise, one has instead,
\be
A(z) = \sum_{i,j} {c_{ij} \over z - z_{ij}} + f(z)
\ee
where $f(z)$ is a holomorphic function of $z$.
When $f(0)$ has to be included to get the correct physical amplitude $A(0)$,
new ways have to be devised to calculate the extra function $f(z)$.

Now we show that by choosing reference lines properly, $A(z)$ vanishes indeed as $z \rightarrow \infty$,
when fermion lines are present.
First, we use the MHV Feynman diagrams. In this case, $A(\infty)$ does vanish if
no two adjacent fermion lines are taken as reference lines, as mentioned in section 3.
Otherwise,  $A(\infty)$ is finite, which can easily be verified.

If we want to make an argument independent of MHV Feynman diagrams, 
one has to constrain the types of reference lines properly, just as in the case of pure gluons.
To see this, we first classify limiting behaviors of vertices and external and internal lines as $z \rightarrow \infty$.
The behaviors of gluon vertices, gluon internal and external lines
are the same as those characterized in section 3.1 of \cite{bcfw}.
Now fermions.
Fermion-fermion-gluon vertices are independent of $z$
and fermion internal lines approach to constants at the worst.
Depending on the helicity of the shifted fermion and the way it is shifted,
an external fermion line would approach a constant or infinity $\sim z$.
So it is possible that one may end up with a wrong $A(z)$ if reference lines are not chosen properly.
For example, when two fermion lines are picked up as reference lines,
$A(z)$ looks as if could approach a constant at the best and an infinity $\sim z^2$ at the worst.
Experience tells that, also supported by MHV Feynman diagrams,
the behavior of $A(z)$ is actually much better,
though the present argument cannot prove cancellations of bad $z$ behaviors in general.
Of course, one can usually stay on the safe side, by choosing ``good" reference lines.
For example, one can choose two external gluons of helicities $(-,+)$ for reference,
$A(z)$ would vanish $\sim 1/z^2$, 
as now all $z$ factors from internal gluons and gluon vertices cancel out, due to the presence of fermion lines.
If one chooses one fermion and one gluon $(\Lambda^-, g^+)$ for reference,
$A(z)$ would vanish $\sim 1/z$ as $z \rightarrow \infty$.

\acknowledgments
We would like to thank C. Berger, L. Dixon, B. Feng, D. Kosower and M. Mangano for correspondences.
This work is supported in part by the National Science Foundation of China (10425525).



\begin{thebibliography}{999}
\bibitem{PT} S. Parke and T. Taylor,
``An Amplitude For N Gluon Scattering",
Phys. Rev. Lett. 56, 2459 (1986); 
F. A. Berends and W. T. Giele,
``Recursive Calculations For Processes With N Gluons",
 Nucl. Phys. B306, 759 (1988).
\bibitem{witten} E. Witten,
``Perturbative Gauge Theory as a String Theory in Twistor Space",
hep-th/0312171.
\bibitem{penrose} R. Penrose, 
``Twistor Algebra",
J. Math. Phys. 8, 345 (1967).
\bibitem{csw} F. Cachazo, P. Svrcek and E. Witten, 
``MHV Vertices and Tree Amplitudes in Gauge Theory",
hep-th/0403047.
\bibitem{wuzhu} J. B.Wu and C. J. Zhu, 
``MHV Vertices and Scattering Amplitudes in Gauge Theory",
hep-th/0406085.
\bibitem{khoze} V. Khoze,
``Gauge Theory Amplitudes, Scalar Graphs And Twistor Space",
hep-th/0408233.
\bibitem{ggk} G. Georgiou, E. W. N. Glover and V. V. Khoze,
``Non-MHV Tree Amplitudes In Gauge Theory",
JHEP 0407:048 (2004), hep-th/0407027. 
\bibitem{kosower} D. A. Kosower,
``Next-To-Maximal Helicity Violating Amplitudes In Gauge Theory",
hep-th/0406175. 
\bibitem{wuzhu2} J. B. Wu, C. J. Zhu,
``MHV Vertices And Fermionic Scattering Amplitudes In Gauge Theory With Quarks And Gluinos",
hep-th/0406146.
\bibitem{bbk} I. Bena, Z. Bern and D. A. Kosower,
``Twistor-Space Recursive Formulation Of Gauge-Theory Amplitudes",
hep-th/0406133.
\bibitem{wuzhu3} J. B. Wu, C. J. Zhu,
``MHV Vertices And Scattering Amplitudes In Gauge Theory",
JHEP 0407:032 (2004), hep-th/0406085.
\bibitem{gk} G. Georgiou, V. V. Khoze,
``Tree Amplitudes In Gauge Theory As Scalar MHV Diagrams",
JHEP 0405:070 (2004), hep-th/0404072.
\bibitem{zhu} C. J. Zhu,
``The Googly Amplitudes In Gauge Theory",
JHEP 0404:032 (2004), hep-th/0403115.
\bibitem{bst} A. Brandhuber, B. Spence and G. Travaglini,
``One-Loop Gauge Theory Amplitudes In N=4 Super Yang-Mills From MHV Vertices",
hep-th/0407214.
\bibitem{lw1} M. Luo and C. Wen,
``One-Loop Maximal Helicity Violating Amplitudes in $N=4$ Super Yang-Mills Theories",
JHEP 0411:004 (2004), hep-th/0410045.
\bibitem{lw2} M. Luo and C. Wen,
``Systematics of One-Loop Scattering Amplitudes in $N=4$ Super Yang-Mills Theories",
hep-th/0410118.
\bibitem{csw2} F. Cachazo, P. Svrcek and E. Witten,
``Gauge Theory Amplitudes In Twistor Space And Holomorphic Anomaly",
hep-th/0409245.
\bibitem{csw3} F. Cachazo, P. Svrcek, E.Witten,
``Twistor Space Structure Of One-Loop Amplitudes In Gauge Theory",
hep-th/0406177.
\bibitem{bern10} I. Bena, Z. Bern, D. A. Kosower, R. Roiban.
``Loops in Twistor Space",
hep-th/0410054.
\bibitem{cachazo10} F. Cachazo,
``Holomorphic Anomaly of Unitary Cuts and One-Loop Gauge Theory Amplitudes",
hep-th/0410077.
\bibitem{rsv1}
R. Roiban, M, Spradlin, and A. Volovich,
``A Googly Amplitude from the B-model in Twistor Space",
JHEP 0404:012 (2004), hep-th/0402016.
\bibitem{rsv2}
R. Roiban, A. Volovich,
``All Googly Amplitudes from the B-model in Twistor Space",
hep-th/0402121.
\bibitem{rsv3}
R. Roiban, M, Spradlin, and A. Volovich,
``On the Tree-Level S-Matrix of Yang-Mills Theory",
Phys.Rev. D70, 026009 (2004), hep-th/0403190.
\bibitem{rsv4} R. Roiban, M. Spradlin, A. Volovich,
``Dissolving N= 4 Loop Amplitudes into QCD Tree Amplitudes",
hep-th/0412265.
\bibitem{bcf} R. Britto, F. Cachazo, B. Feng,
``New Recursion Relations for Tree Amplitudes of Gluons",
hep-th/0412308.
\bibitem{bcfw} R. Britto, F. Cachazo, B. Feng, E. Witten,
``Direct Proof of Tree-Level Recursion Relation in Yang-Mills Theory",
hep-th/0501052.
\bibitem{bk1} Z. Bern, V. Del Duca, L. Dixon, and D. A. Kosower,
``All Non-Maximally-Helicity-Violating One-Loop Seven-Gluon Amplitudes in N=4 Super-Yang-Mills Theory",
hep-th/0410224.
\bibitem{bk2} Z. Bern, L. Dixon, and D. A. Kosower,
``All Next-to-Maximally-Helicity-Violating One-Loop Gluon Amplitudes in N=4 Super-Yang-Mills Theory",
hep-th/0412210.
\bibitem{mp} M. L. Mangano, S. J. Parke,
``Multiparton Amplitudes in Gauge Theories",
Phys.Rept.200, 301 (1991).
\bibitem{kosower1} D. A. Kosower,
``Light Cone Recurrence Relations for QCD Amplitudes",
Nucl. Phys. B335:23 (1990).
\end{thebibliography}
\end{document}